\documentclass[notitlepage,nofootinbib,preprintnumbers,amssymb,superscriptaddress]{revtex4-2}
\usepackage{amsfonts,amssymb,mathtools,graphicx,color,bm}
\usepackage{autobreak}
\definecolor{ultramarine}{rgb}{0.07, 0.04, 0.56}
\definecolor{cadmiumgreen}{rgb}{0.0, 0.42, 0.24}
\definecolor{indigo(dye)}{rgb}{0.0, 0.25, 0.42}
\usepackage[linktocpage=true]{hyperref}
\hypersetup{
colorlinks=true,
citecolor=ultramarine,
linkcolor=cadmiumgreen,
urlcolor=indigo(dye),
}

\let\originalleft\left
\let\originalright\right
\renewcommand{\left}{\mathopen{}\mathclose\bgroup\originalleft}
\renewcommand{\right}{\aftergroup\egroup\originalright}
\newcommand{\D}{{\rm d}}

\newcommand{\bra}[1]{\left( #1 \right)}  
\newcommand{\brb}[1]{\left[ #1 \right]}  
\newcommand{\brc}[1]{\left\{ #1 \right\}}  
\newcommand{\be}{\begin{equation}}  
\newcommand{\ee}{\end{equation}}
\newcommand{\bem}{\begin{bmatrix}}
\newcommand{\eem}{\end{bmatrix}}

\newcommand{\mL}{\mathcal{L}}

\begin{document}

\preprint{RESCEU-29/26}

\title{Inverse design of scalar-tensor EFT from odd-parity effective potentials}

\author{Kazufumi Takahashi}
\affiliation{Research Center for the Early Universe (RESCEU), Graduate School of Science, The University of Tokyo, Tokyo 113-0033, Japan}

\author{Yuki Watanabe}
\affiliation{National Institute of Technology, Gunma College, Maebashi, Gunma 371-8530, Japan}
\affiliation{Research Center for the Early Universe (RESCEU), Graduate School of Science, The University of Tokyo, Tokyo 113-0033, Japan}

\begin{abstract}
We develop an inverse-design approach to the effective field theory (EFT) of perturbations in scalar-tensor gravity around static and spherically symmetric backgrounds with a timelike scalar-field gradient, taking the odd-parity effective potential as input and reconstructing compatible background and EFT data. As analytically tractable examples, we construct explicit realizations of the P{\"o}schl-Teller potential, for which the quasinormal-mode spectrum can be obtained in closed form. We then study the Regge-Wheeler potential of general relativity and its parametrized deformations, showing how both can be realized within the scalar-tensor EFT. In particular, the latter realizations provide a route to interpreting phenomenological constraints on potential deformations from black hole ringdown observations in terms of the underlying EFT data.
\end{abstract}

\maketitle

\section{Introduction}\label{sec:intro}

The advent of gravitational-wave astronomy has opened a new observational window onto the dynamical and strong-field regime of gravity~\cite{Abbott:2016blz}.
Since the first detection, the number of observed compact-binary coalescence signals has increased rapidly, and the expanding catalog of gravitational-wave transients provides an increasingly rich data set for probing strong-field gravity~\cite{LIGOScientific:2026sit}.
For binary black hole (BH) coalescences, the post-merger ringdown is characterized by damped oscillations of the remnant BH associated with its quasinormal modes~\cite{Berti:2009kk,Konoplya:2011qq,Berti:2025hly}.
Their characteristic frequencies and damping times encode the dynamics of linear perturbations around the remnant spacetime and can therefore provide a probe of the underlying gravitational theory.
To date, tests of general relativity (GR) using gravitational-wave observations have found no compelling evidence for departures from the predictions of GR~\cite{LIGOScientific:2026oim}.

Despite the current agreement with GR, it remains important to develop theoretical frameworks that allow possible deviations from GR to be studied in a systematic manner.
Such comparisons are valuable regardless of their outcome: a discrepancy would point to new gravitational dynamics, while continued agreement would further strengthen the empirical support for GR.
One of the simplest and most widely studied extensions of GR is scalar-tensor gravity, in which an additional scalar field is introduced alongside the spacetime metric.
Extensive efforts have been devoted to constructing general classes of scalar-tensor theories, including Horndeski theories~\cite{Horndeski:1974wa,Deffayet:2011gz,Kobayashi:2011nu}, degenerate higher-order scalar-tensor (DHOST) theories~\cite{Langlois:2015cwa,Crisostomi:2016czh,BenAchour:2016fzp,Takahashi:2017pje,Langlois:2018jdg}, U-DHOST theories~\cite{DeFelice:2018ewo,DeFelice:2021hps}, generalized disformal theories~\cite{Takahashi:2021ttd,Takahashi:2022mew,Takahashi:2023jro,Takahashi:2023vva}, and more recent higher-derivative extensions~\cite{Michiwaki:2026xru,Gorji:2026nes}.

A particularly useful way to study perturbations around a given background in broad classes of such theories is the effective field theory (EFT) approach.
Historically, the EFT approach to cosmological perturbations was developed through studies of ghost condensation~\cite{ArkaniHamed:2003uy,Arkani-Hamed:2003juy}, inflation~\cite{Cheung:2007st}, and dark energy~\cite{Gubitosi:2012hu}.
The key idea is to organize the action for perturbations according to the symmetries of the background.
For a homogeneous and isotropic cosmological background with a timelike scalar-field gradient, the scalar profile defines a preferred time slicing, and in unitary gauge, where the scalar field is used as the time coordinate, the action can be constructed from quantities invariant under the residual spatial diffeomorphisms.
This makes it possible to describe a broad class of scalar-tensor theories without specifying the details of the underlying covariant Lagrangian.
More recently, the EFT construction was generalized to arbitrary inhomogeneous backgrounds while retaining a timelike scalar-field gradient~\cite{Mukohyama:2022enj}.\footnote{For an analogous EFT construction with a spacelike scalar gradient, see Ref.~\cite{Mukohyama:2025jzk}. The EFT framework has also been extended to vector-tensor gravity. Building on gauged ghost condensation on Minkowski spacetime~\cite{Cheng:2006us,Mukohyama:2006mm}, the construction was generalized first to cosmological backgrounds~\cite{Aoki:2021wew} and subsequently to arbitrary backgrounds with a timelike vector profile~\cite{Aoki:2023bmz}. The latter framework has also been applied to the study of BH quasinormal modes~\cite{Tomizuka:2025dpy}.}
For static and spherically symmetric BH backgrounds, the odd-parity perturbation theory was developed in Refs.~\cite{Mukohyama:2022skk,Khoury:2022zor,Mukohyama:2023xyf}.
In particular, the master equation for odd-parity perturbations takes the form of a generalized Regge-Wheeler equation~\cite{Mukohyama:2022skk,Mukohyama:2023xyf}.
The framework has been applied to BH quasinormal modes~\cite{Mukohyama:2023xyf}, tidal Love numbers~\cite{Barura:2024uog}, and greybody factors~\cite{Oshita:2024fzf}.
Spherically symmetric perturbations have also been investigated in Ref.~\cite{Mukohyama:2025owu}.

Most applications of the EFT framework proceed in the forward direction: one specifies a background solution and EFT coefficients, derives the corresponding perturbation equations and effective potentials, and then predicts physical observables.
On the other hand, phenomenological approaches often parametrize possible deviations directly at the level of the perturbation potential and study their consequences for observables, including the quasinormal-mode spectrum~\cite{Cardoso:2019mqo,McManus:2019ulj}.
A related inverse problem is to infer the effective potential from a given quasinormal-mode spectrum, which has also been explored in the literature~\cite{Franciolini:2018uyq,Volkel:2022khh}.
A further complementary inverse problem is to ask which background and EFT data can give rise to a prescribed effective potential.
Such an inverse-design perspective provides a systematic bridge from phenomenological modifications of the perturbation potential toward the underlying EFT description, thereby providing an intermediate step toward reconstructing an underlying spacetime-covariant scalar-tensor theory from perturbation dynamics.
It also makes explicit the possible degeneracy of the reconstruction, since a given effective potential can in general admit multiple EFT realizations.

Motivated by this perspective, in the present paper, we develop an inverse-design approach and illustrate it with several examples.
First, we consider two analytically tractable realizations with a P\"{o}schl-Teller potential, one admitting a Schwarzschild-de Sitter-type static patch and the other describing a wormhole-type static patch.
We then consider the Regge-Wheeler potential and characterize EFT realizations that reproduce the GR potential exactly, showing that the same potential can arise from a large family of realizations distinct from the standard GR configuration.
Finally, we consider parametrized deformations of the Regge-Wheeler potential and construct explicit realizations of such phenomenological modifications.

The rest of this paper is organized as follows.
In Sec.~\ref{sec:model}, we review the EFT framework for odd-parity perturbations and derive the generalized Regge-Wheeler equation on a broad class of static and spherically symmetric backgrounds.
In Sec.~\ref{sec:inverse_design}, we formulate the inverse-design problem and present realizations of the P\"{o}schl-Teller and Regge-Wheeler potentials, together with parametrized deformations of the latter.
Finally, we draw our conclusions in Sec.~\ref{sec:conc}.

\section{EFT framework for odd-parity perturbations}\label{sec:model}

In this section, we briefly review the EFT of perturbations around backgrounds with a timelike scalar-field gradient developed in Refs.~\cite{Mukohyama:2022enj,Mukohyama:2022skk,Mukohyama:2023xyf} and extend the odd-parity analysis to a broader class of static and spherically symmetric backgrounds relevant to our inverse-design problem, deriving the corresponding generalized Regge-Wheeler equation.

\subsection{EFT action}\label{ssec:EFT_action}

Let $\phi$ be a scalar field whose gradient is timelike.
We define the unit normal to the constant-$\phi$ hypersurfaces and the scalar kinetic term by
    \begin{align}
    n_\mu\coloneqq-\frac{\partial_\mu\phi}{\sqrt{-X}}, \qquad
    X\coloneqq g^{\mu\nu}\partial_\mu\phi\,\partial_\nu\phi<0,
    \label{eq:normal_EFT}
    \end{align}
where the overall sign of $n^\mu$ is chosen so that $n^\mu$ is future-directed.
The induced metric and the extrinsic curvature are
    \begin{align}
    h_{\mu\nu}\coloneqq g_{\mu\nu}+n_\mu n_\nu, \qquad
    K_{\mu\nu}\coloneqq h_\mu^{\lambda}\nabla_\lambda n_\nu.
    \end{align}
In unitary gauge, the scalar field is homogeneous on constant-time hypersurfaces, so that the scalar fluctuation is absorbed into the metric and the residual gauge symmetry is the spatial diffeomorphism invariance.
We choose the unitary-gauge time coordinate~$\tau$ such that $\phi=q\tau$, where $q>0$ is a constant.
Accordingly, the EFT action can be constructed from quantities covariant under these spatial diffeomorphisms, including quantities invariant under the full spacetime diffeomorphisms, schematically as
    \begin{align}
    S=\int \D^4x\sqrt{-g}\,\mL\bra{g^{\tau\tau},K^\mu_\nu,{}^{(3)}\!R^\mu_\nu,\nabla_\mu},
    \label{eq:EFT_general}
    \end{align}
where ${}^{(3)}\!R^\mu_\nu$ is the three-dimensional Ricci tensor associated with the induced metric~$h_{\mu\nu}$.
It should be noted that the spacetime curvature is not an independent building block because of the Gauss-Codazzi equations.
Note also that the Lagrangian can in general depend explicitly on $\tau$.
In what follows, however, we restrict ourselves to theories that are shift-symmetric with respect to $\phi$, for which the Lagrangian has no explicit dependence on $\tau$.

In practice, one expands the EFT action~\eqref{eq:EFT_general} in perturbations about a given background and organizes the resulting operators in a derivative expansion.
At fixed orders in perturbations and derivatives, the action is characterized by a finite number of EFT coefficient functions.
For an inhomogeneous background, we define the perturbations as
    \begin{align}
    \delta g^{\tau\tau}\coloneqq g^{\tau\tau}-\bar{g}^{\tau\tau}(\tau,x^i), \qquad
    \delta K^\mu_\nu\coloneqq K^\mu_\nu-\bar{K}^\mu_\nu(\tau,x^i), \qquad
    \delta{}^{(3)}\!R^\mu_\nu\coloneqq{}^{(3)}\!R^\mu_\nu-{}^{(3)}\!\bar{R}^\mu_\nu(\tau,x^i),
    \label{eq:EFT_pert}
    \end{align}
where a bar denotes the background value, which may depend on both $\tau$ and the spatial coordinates~$x^i$.
Accordingly, the Taylor coefficients appearing in the perturbative expansion of Eq.~\eqref{eq:EFT_general} can depend on the spatial coordinates.
However, such spatial dependence is not arbitrary: the coefficients satisfy consistency relations that ensure that the expanded action retains the spatial diffeomorphism invariance of the original action~\cite{Mukohyama:2022enj,Mukohyama:2022skk}.
Also, following Refs.~\cite{Mukohyama:2022enj,Mukohyama:2022skk}, we use the curvature combination
    \begin{align}
    R\coloneqq{}^{(3)}\!R+K_{\mu\nu}K^{\mu\nu}-K^2
    =\tilde{R}-2\nabla_\mu\bra{Kn^\mu-n^\nu\nabla_\nu n^\mu},
    \label{eq:EFT_R}
    \end{align}
where $\tilde{R}$ is the ordinary four-dimensional Ricci scalar.

In the present paper, we mainly focus on odd-parity perturbations around a static and spherically symmetric background with a timelike scalar-field gradient, $\phi=q\tau$.
Accordingly, we restrict ourselves to a subset of EFT operators that describes the background equations and the odd-parity dynamics considered below.\footnote{For simplicity, we do not include the additional operator~$M_6(r)\bar{\sigma}^\mu_\nu\delta K^\nu_\alpha\delta K^\alpha_\mu$, where $\bar{\sigma}^\mu_\nu$ denotes the traceless part of $\bar{K}^\mu_\nu$. This operator can also contribute to odd-parity perturbations on an inhomogeneous background and, in general, leads to different propagation speeds in the radial and angular directions~\cite{Mukohyama:2024pqe}.}
In particular, we consider the following truncated EFT action:
    \begin{align}
    S_{\rm odd}=\int \D^4x\sqrt{-g}\brb{\frac{M^2(r)}{2}R-\Lambda(r)-c(r)g^{\tau\tau}-\tilde{\beta}(r)K-\alpha(r)\bar{K}^\mu_\nu K^\nu_\mu
    -\zeta(r)\bar{n}^\mu\partial_\mu g^{\tau\tau}+\frac{M_3^2(r)}{2}\delta K^\mu_\nu\delta K^\nu_\mu}.
    \label{EFT_action}
    \end{align}
Despite the notation, $M^2$ and $M_3^2$ denote independent EFT coefficients of mass dimension two and are not assumed to be positive at this stage.
Owing to the static and spherical symmetry of the background, together with the shift symmetry with respect to $\phi$, the EFT coefficients can be taken to depend only on a radial coordinate~$r$, with no explicit dependence on $\tau$.
Operators that contribute only to the even-parity sector at quadratic order, such as $(\delta g^{\tau\tau})^2$ and $(\delta K)^2$, are not displayed.
Note that a given spacetime-covariant scalar-tensor theory maps onto a specific set of EFT parameters.
Explicit dictionaries for Horndeski theories and shift- and reflection-symmetric quadratic (D)HOST theories were derived in Refs.~\cite{Mukohyama:2022enj,Mukohyama:2022skk}, respectively, with $\bar{g}^{\tau\tau}=-1$ assumed in the latter, implying $\bar{X}={\rm const}$.

It is useful at this point to make explicit how the consistency relations implied by spatial diffeomorphism invariance constrain the coefficients retained in Eq.~\eqref{EFT_action}.
At the derivative order considered here, the relation involving only the coefficients appearing at the tadpole level takes the form
    \begin{align}
    \Lambda'+\bar{g}^{\tau\tau}c'-\frac{1}{2}\bar{R}(M^2)'+\bar{K}\tilde{\beta}'
    +\bar{K}^\mu_\nu\bar{K}^\nu_\mu\alpha'
    +\frac{\alpha}{2}\bra{\bar{K}^\mu_\nu\bar{K}^\nu_\mu}'
    +(\zeta\bar{n}^\mu)'\partial_\mu\bar{g}^{\tau\tau}=0,
    \label{eq:EFT_consistency_tadpole}
    \end{align}
where a prime denotes differentiation with respect to $r$.
Since all the EFT coefficients depend only on $r$ in the present setup, only the radial component of the corresponding spatial consistency relation is nontrivial, while the angular components are identically satisfied.
The remaining consistency relations also involve coefficients of quadratic operators that do not affect the odd-parity dynamics considered below.
Since such coefficients are otherwise irrelevant to the present analysis, we do not display those relations explicitly and assume that the omitted coefficients are chosen so that they are satisfied.

In the EFT action~\eqref{EFT_action}, the terms linear in perturbations yield the background equations, or equivalently the tadpole cancellation conditions.
They can be written as
    \begin{align}
    \hat{\cal{E}}_{\mu\nu}M^2=\bar{T}_{\mu\nu},
    \label{eq:tadpole_general}
    \end{align}
where
    \begin{align}
    \hat{\cal{E}}_{\mu\nu}\coloneqq
    \bar{G}_{\mu\nu}
    -\bar{\rm D}_\mu\bar{\rm D}_\nu
    +\bar{h}_{\mu\nu}\bar{\rm D}_\lambda\bar{\rm D}^\lambda
    +2\bra{\bar{h}_{\mu\nu}\bar{a}^\lambda-\bar{a}_{(\mu}\bar{h}_{\nu)}^\lambda}\bar{\rm D}_\lambda
    +\bra{\bar{K}_{\mu\nu}-\bar{K}\bar{h}_{\mu\nu}}\bar{n}^\lambda\bar{\nabla}_\lambda
    -2\bar{n}_{(\mu}\bra{\bar{K}_{\nu)}^\lambda-\bar{K}\bar{h}_{\nu)}^\lambda}\bar{\rm D}_\lambda.
    \label{eq:tadpole_operator}
    \end{align}
Here, $\rm D_\mu$ denotes the covariant derivative compatible with the induced metric~$h_{\mu\nu}$ on the constant-$\phi$ hypersurfaces, and $a_\mu\coloneqq n^\nu\nabla_\nu n_\mu$.
The effective background stress-energy tensor on the right-hand side of Eq.~\eqref{eq:tadpole_general} is
    \begin{align}
    \bar{T}_{\mu\nu}&=-\bra{
    \Lambda+c\bar{g}^{\tau\tau}
    +\zeta\bar{n}^\lambda\partial_\lambda\bar{g}^{\tau\tau}
    -\bar{n}^\lambda\partial_\lambda\tilde{\beta}
    +\alpha\bar{K}^\lambda_\sigma\bar{K}^\sigma_\lambda
    }\bar{g}_{\mu\nu}
    +\brb{
    -2c\bar{g}^{\tau\tau}
    -\bar{n}^\lambda\partial_\lambda\tilde{\beta}
    +\alpha\bar{K}^\lambda_\sigma\bar{K}^\sigma_\lambda
    +2\bar{g}^{\tau\tau}\bar{\nabla}_\lambda\bra{\zeta\bar{n}^\lambda}
    }\bar{n}_\mu\bar{n}_\nu
    \nonumber\\
    &\quad -2\bar{n}_{(\mu}\partial_{\nu)}\tilde{\beta}
    +2\alpha\bar{K}_\mu^\lambda\bar{K}_{\lambda\nu}
    -2\bar{\nabla}_\lambda\bra{\alpha\bar{K}^\lambda_{(\mu}\bar{n}_{\nu)}}
    +\bar{\nabla}_\lambda\bra{\alpha\bar{K}_{\mu\nu}\bar{n}^\lambda}.
    \label{eq:Ttadpole}
    \end{align}
For the static and spherically symmetric background introduced below, Eq.~\eqref{eq:tadpole_general} gives four independent background equations, which will be used as reconstruction equations for the tadpole coefficients.

\subsection{Odd-parity perturbations around a static and spherically symmetric background}\label{ssec:odd_parity}

We take the background metric and scalar field to be
    \begin{align}
    \bar{g}_{\mu\nu}\D x^\mu\D x^\nu
    =-A(r)\D t^2+\frac{\D r^2}{B(r)}+{\cal R}(r)^2\D\Omega^2, \qquad
    \bar{\phi}=qt+\psi(r),
    \label{eq:metric_BG}
    \end{align}
where $\D\Omega^2=\D\theta^2+\sin^2\theta\,\D\varphi^2$ denotes the line element on the unit two-sphere and ${\cal R}(r)$ is the areal radius.
In shift-symmetric scalar-tensor theories, such a linearly time-dependent scalar profile can be compatible with a static background metric~\cite{Babichev:2013cya}, and perturbations about such BH solutions have been studied extensively in Horndeski, DHOST, and more general higher-order scalar-tensor theories~\cite{Ogawa:2015pea,Takahashi:2016dnv,Tretyakova:2017lyg,Babichev:2018uiw,Takahashi:2019oxz,deRham:2019gha,Khoury:2020aya,Tomikawa:2021pca,Langlois:2021xzq,Langlois:2021aji,Takahashi:2021bml,Nakashi:2022wdg,Nakashi:2023vul,Kobayashi:2025evr}.
We allow for configurations in which ${\cal R}'(r)$ vanishes at some radius, as occurs, e.g., at a wormhole throat.
The scalar kinetic term on this background is
    \begin{align}
    \bar{X}=-\frac{q^2}{A}+B\psi'^2, \qquad
    {\cal X}(r)\coloneqq-\frac{\bar{X}}{q^2}>0,
    \label{eq:Xcal}
    \end{align}
where we have introduced the dimensionless quantity~${\cal X}$ for later convenience.
Here, Eq.~\eqref{eq:Xcal} yields
    \begin{align}
    1-{\cal X}A=\frac{AB\psi'^2}{q^2}\ge 0,
    \label{eq:1-XA}
    \end{align}
provided that $AB>0$.
We introduce the unitary-gauge time coordinate~$\tau$ through
    \begin{align}
    \D\tau=\D t+\frac{\psi'}{q}\D r
    =\D t+\frac{\sqrt{\cal X}{\cal Z}}{A{\cal Y}}\D r,
    \label{eq:unitary_coord}
    \end{align}
where we have defined
    \begin{align}
    {\cal Y}\coloneqq\sqrt{\frac{B}{A}}, \qquad
    {\cal Z}\coloneqq \frac{A{\cal Y}\psi'}{q\sqrt{\cal X}}
    =\pm\sqrt{\frac{1}{\cal X}-A}.
    \label{eq:YZ_def}
    \end{align}
Here, ${\cal Z}$ is allowed to take either sign, thereby keeping track of the sign of $\psi'$, and may change sign smoothly where $\psi'$ vanishes.
In the $(\tau,r)$ coordinates, the background metric and scalar field take the form
    \begin{align}
    \bar{g}_{\mu\nu}\D x^\mu\D x^\nu
    =-A\,\D\tau^2+\frac{2\sqrt{\cal X}{\cal Z}}{\cal Y}\D\tau\D r+\frac{\cal X}{{\cal Y}^2}\D r^2+{\cal R}^2\D\Omega^2, \qquad
    \bar{\phi}=q\tau.
    \label{eq:metric_unitary}
    \end{align}
It should be noted that $\bar{g}^{\tau\tau}=-{\cal X}$.

On the background defined above, the tadpole cancellation condition~\eqref{eq:tadpole_general} yields four independent equations.
After some manipulations, three of them can be written as
    \begin{align}
    \begin{split}
    \Lambda&=-\frac{\cal Y}{{\cal R}^2}\bra{{\cal R}^2{\cal X}{\cal Z}\zeta}'-\frac{{\cal Y}({\cal R}'{\cal Y}M^2)'}{{\cal R}{\cal X}}+\frac{M^2}{{\cal R}^2}\brb{1-{\cal R}'{\cal Y}^2(A{\cal R})'}, \\
    c&=-\frac{\cal Y}{{\cal R}^2}\bra{{\cal R}^2{\cal Z}\zeta}'-\frac{{\cal Y}M^4}{{\cal R}{\cal X}^2}\bra{\frac{{\cal R}'{\cal Y}}{M^2}}', \\
    \tilde{\beta}'&=\frac{1}{{\cal R}^2}\brb{\bra{{\cal R}^2{\cal Y}{\cal Z}'\alpha}'-2{\cal R}'^2{\cal Y}{\cal Z}\alpha}+\frac{2{\cal Z}({\cal R}'{\cal Y}M^2)'}{\cal R},
    \end{split}
    \label{eq:tadpole_spherical_1}
    \end{align}
while the remaining equation takes the form
    \begin{align}
    \brb{{\cal R}^3{\cal Y}{\cal Z}\bra{\frac{\cal Z}{{\cal R}}}'\alpha}'
    =-\frac{M^2}{\cal Y}-\frac{2{\cal R}({\cal R}'{\cal Y})'M^2}{\cal X}-\frac{1}{2}\brb{{\cal R}^4{\cal Y}\bra{\frac{A}{{\cal R}^2}}'M^2}'-{\cal R}^2\brb{\frac{\cal Y}{\cal X}(M^2)'}'.
    \label{eq:tadpole_spherical_2}
    \end{align}
Note that we have assumed $M^2\ne 0$, which is justified since the stability of odd-parity perturbations requires $M^2>0$ [see Eq.~\eqref{eq:odd_health}].
It should also be noted that the tadpole cancellation conditions determine only $\tilde{\beta}'$, rather than $\tilde{\beta}$ itself.
This is expected since the term~$-\tilde{\beta}K$ in the EFT action~\eqref{EFT_action} is equivalent to $n^\mu\partial_\mu\tilde{\beta}$ up to a total derivative.
Hence, the integration constant arising when solving for $\tilde{\beta}$ does not affect the equations of motion.

Upon using the tadpole cancellation conditions above, the consistency relation~\eqref{eq:EFT_consistency_tadpole} reduces to
    \begin{align}
    \bra{\frac{{\cal R}^2{\cal X}'{\cal Y}{\cal Z}^2}{\sqrt{\cal X}}\zeta}'=0.
    \label{eq:consistency_spherical}
    \end{align}
For ${\cal X}'=0$, this condition is identically satisfied, as in the setups considered in Refs.~\cite{Mukohyama:2022skk,Mukohyama:2023xyf}.
For a radially varying~${\cal X}$, on the other hand, it imposes an additional constraint on $\zeta$.

Let us now study odd-parity perturbations about the above background.
In general, the odd-parity metric perturbations are expanded in terms of the spherical harmonics as
    \begin{align}
    \begin{split}
    \delta g_{\tau\tau}&=\delta g_{\tau r}=\delta g_{rr}=0, \\
    \delta g_{\tau a}&=\sum_{\ell,m}{\cal R}^2h_{0,\ell m}(\tau,r)E_a{}^{b}\hat{\nabla}_bY_{\ell m}(\theta,\varphi), \\
    \delta g_{ra}&=\sum_{\ell,m}{\cal R}^2h_{1,\ell m}(\tau,r)E_a{}^{b}\hat{\nabla}_bY_{\ell m}(\theta,\varphi), \\
    \delta g_{ab}&=\sum_{\ell,m}{\cal R}^2h_{2,\ell m}(\tau,r)E_{(a|}{}^{c}\hat{\nabla}_c\hat{\nabla}_{|b)}Y_{\ell m}(\theta,\varphi),
    \end{split} \label{eq:odd_decomp}
    \end{align}
where $a,b\in\{\theta,\varphi\}$, $E_{ab}$ is the antisymmetric tensor on the unit two-sphere, and $\hat{\nabla}_a$ is the covariant derivative compatible with the unit two-sphere metric.
The function~$h_2$ is absent for the dipole mode~$\ell=1$.
Note that there is a gauge degree of freedom associated with an infinitesimal odd-parity spatial diffeomorphism~$x^\mu\to x^\mu+\epsilon^\mu$, parametrized by
    \begin{align}
    \epsilon^\tau=\epsilon^r=0, \qquad
    \epsilon^a=\sum_{\ell,m}\Xi_{\ell m}(\tau,r)E^{ab}\hat{\nabla}_bY_{\ell m}(\theta,\varphi).
    \end{align}
For each mode, the perturbation variables transform as
    \begin{align}
    h_{0,\ell m}\to h_{0,\ell m}-\dot{\Xi}_{\ell m}, \qquad
    h_{1,\ell m}\to h_{1,\ell m}-\Xi_{\ell m}', \qquad
    h_{2,\ell m}\to h_{2,\ell m}-2\Xi_{\ell m},
    \label{eq:odd_gauge}
    \end{align}
where a dot denotes differentiation with respect to $\tau$.
Thanks to the spherical symmetry of the background, it suffices to consider the modes with $m=0$.
Accordingly, we use the Legendre polynomials~$P_\ell(\cos\theta)$ in place of $Y_{\ell0}$, with the normalization absorbed into the perturbation variables.
Moreover, modes with different values of $\ell$ are decoupled.
Therefore, in what follows, we omit the indices~$\ell$ and $m$.
Although the dipole mode~$\ell=1$ carries no propagating degree of freedom, it describes a slowly rotating deformation of the background.
For $\ell\geq2$, we impose the Regge-Wheeler gauge~$h_2=0$, which completely fixes the odd-parity gauge freedom and hence can be imposed at the action level~\cite{Motohashi:2016prk}.
Note that the following odd-parity analysis is insensitive to the scalar-field perturbation, since it belongs to the even-parity sector.
Therefore, the analysis also applies to theories in which the scalar degree of freedom does not propagate, such as cuscuton~\cite{Afshordi:2006ad} and its extension~\cite{Iyonaga:2018vnu}, as well as spatially covariant theories with only two tensorial degrees of freedom~\cite{Gao:2019twq}.

Substituting Eq.~\eqref{eq:odd_decomp} with $h_2=0$ into the EFT action, integrating over the angular variables, and using the tadpole cancellation conditions~\eqref{eq:tadpole_spherical_1} and \eqref{eq:tadpole_spherical_2} yields a quadratic Lagrangian for $h_0$ and $h_1$, which can be written as
    \begin{align}
    \mL_{\rm odd}^{(2)}=p_0h_0^2+p_1h_1^2+2p_2h_0h_1
    +p_3\left[\bra{\dot{h}_1-h_0'}^2+2p_4h_0'h_1\right],
    \label{eq:odd_quad_general}
    \end{align}
up to total derivatives and an irrelevant overall positive constant.
Here, the coefficients~$p_i$ are given by
    \begin{align}
    \begin{split}
    &p_0=\frac{(\ell-1)(\ell+2){\cal R}^2{\cal X}}{2{\cal Y}}(M^2+M_3^2), \qquad
    p_1=-\frac{(\ell-1)(\ell+2){\cal R}^2{\cal Y}}{2{\cal X}}\brb{M^2-{\cal X}{\cal Z}^2(M^2+M_3^2)}, \\
    &p_2=-\frac{(\ell-1)(\ell+2){\cal R}^2\sqrt{\cal X}{\cal Z}}{2}(M^2+M_3^2), \qquad
    p_3=\frac{{\cal R}^4{\cal Y}}{2}(M^2+M_3^2), \qquad
    p_4=-\frac{{\cal R}{\cal Y}}{\sqrt{\cal X}}\bra{\frac{\cal Z}{\cal R}}'\frac{M_3^2+\alpha}{M^2+M_3^2}.
    \end{split}\label{eq:pi}
    \end{align}
Compared with Ref.~\cite{Mukohyama:2022skk}, we use the generally non-diagonal unitary-gauge coordinates~$(\tau,r)$ rather than the diagonal unitary-gauge coordinates employed there.
Accordingly, an algebraic mixing term proportional to $h_0h_1$ is present in Eq.~\eqref{eq:odd_quad_general}, while $p_4$, which controls the $h_0'h_1$ term, is still proportional to the same combination~$M_3^2+\alpha$.

In what follows, we impose
    \begin{align}
    M_3^2+\alpha=0,
    \label{eq:p4_zero}
    \end{align}
so that $p_4=0$.
The motivation for this restriction comes from the odd-parity dynamics in the presence of a nonvanishing~$p_4$.
As discussed in Ref.~\cite{Mukohyama:2022skk}, without imposing Eq.~\eqref{eq:p4_zero}, one encounters either a divergent propagation speed of the odd mode at spatial infinity or the absence of a nontrivial dipole solution corresponding to slow rotation.
We have verified that the same conclusion continues to hold for asymptotically flat backgrounds within the more general setup considered here.
Moreover, Eq.~\eqref{eq:p4_zero} is automatically satisfied by the shift-symmetric quadratic higher-order scalar-tensor theories covered by the EFT~\cite{Takahashi:2019oxz,Tomikawa:2021pca,Mukohyama:2022skk}.

Once Eq.~\eqref{eq:p4_zero} is imposed, all derivatives of the perturbation variables enter the quadratic Lagrangian through the combination~$\dot{h}_1-h_0'$.
Introducing an auxiliary variable~$\tilde{h}$ for this combination renders both $h_0$ and $h_1$ algebraic, so that they can be integrated out to obtain a quadratic action for the single variable~$\tilde{h}$ for each $\ell\geq2$.\footnote{More explicitly, introducing $\tilde{h}$ through $p_3(\dot{h}_1-h_0')^2\to p_3[2\tilde{h}(\dot{h}_1-h_0')-\tilde{h}^2]$, the original perturbation variables are reconstructed as
    \begin{align*}
    h_0&=-\frac{p_2p_3\dot{\tilde{h}}+p_1(p_3\tilde{h})'}{p_0p_1-p_2^2}, \qquad
    h_1=\frac{p_0p_3\dot{\tilde{h}}+p_2(p_3\tilde{h})'}{p_0p_1-p_2^2},
    \end{align*}
where the $p_i$ are defined in Eq.~\eqref{eq:pi}, and the condition~\eqref{eq:p4_zero} has been imposed. Under the stability conditions~\eqref{eq:odd_health}, the denominator is nonvanishing, since $p_0p_1-p_2^2=-(\ell-1)^2(\ell+2)^2{\cal R}^4M^2(M^2+M_3^2)/4$.}
As in Ref.~\cite{Mukohyama:2022skk}, the local stability conditions can be obtained as
    \begin{align}
    M^2(r)>0, \qquad
    M^2(r)+M_3^2(r)>0,
    \label{eq:odd_health}
    \end{align}
which ensure the positivity of the kinetic-term coefficient and the absence of gradient instabilities.
Under the condition~\eqref{eq:p4_zero}, the radial and angular propagation speeds coincide, and Eq.~\eqref{eq:odd_health} ensures that their common squared speed is positive.
We assume that these conditions hold in the region of interest and define
    \begin{align}
    M(r)\coloneqq\sqrt{M^2(r)}, \qquad
    \alpha_T(r)\coloneqq-\frac{M_3^2(r)}{M^2(r)+M_3^2(r)}.
    \label{eq:alphaT_def}
    \end{align}
The quantity~$1+\alpha_T=M^2/(M^2+M_3^2)$ represents this common squared propagation speed.

In the quadratic action for $\tilde{h}$, there is a mixed time-radial derivative, which we eliminate by introducing the shifted time coordinate~$\tilde{t}$ as
    \begin{align}
    \tilde{t}\coloneqq \tau-\int\frac{{\cal X}^{3/2}{\cal Z}}{{\cal Y}({\cal X}A+\alpha_T)}\D r
    =t+\int\frac{\sqrt{\cal X}{\cal Z}\alpha_T}{A{\cal Y}({\cal X}A+\alpha_T)}\D r.
    \label{eq:master_coords}
    \end{align}
At the same time, it is convenient to introduce the generalized tortoise coordinate~$r_*$ as
    \begin{align}r_*\coloneqq\int\frac{\D r}{F}, \qquad
    F(r)\coloneqq\sqrt{\frac{B}{A}}\frac{{\cal X}A+\alpha_T}{{\cal X}\sqrt{1+\alpha_T}}.
    \label{eq:F_def}
    \end{align}
The function~$F$ determines the relation between $r$ and the generalized tortoise coordinate $r_*$.
In particular, a simple zero of $F$ at a finite radius is mapped to $|r_*|\to\infty$ and therefore corresponds to an odd-mode characteristic horizon.
Note that the location of an odd-mode characteristic horizon need not coincide with that of a horizon of the background metric in general~\cite{Tomikawa:2021pca,Langlois:2022ulw,Nakashi:2023vul}.
After these coordinate transformations, the equation for $\tilde{h}$ generally contains a first-derivative term with respect to $r_*$, which can be eliminated by defining the master variable
    \begin{align}
    \Psi\coloneqq 
    \sqrt{\frac{B}{A}}\frac{M{\cal R}^3}{(1+\alpha_T)^{3/4}}\tilde{h}.
    \label{eq:Psi_def}
    \end{align}
The resulting equation of motion then takes the generalized Regge-Wheeler form
    \begin{align}
    \frac{\partial^2\Psi}{\partial r_*^2}
    -\frac{\partial^2\Psi}{\partial\tilde{t}^2}
    -V_{\rm eff}(r)\Psi=0,
    \label{eq:master}
    \end{align}
with the effective potential given by
    \begin{align}
    V_{\rm eff}(r)
    =\sqrt{1+\alpha_T}\,F\brc{\sqrt{\frac{A}{B}}\frac{(\ell-1)(\ell+2)}{{\cal R}^2}
    +\frac{M{\cal R}}{(1+\alpha_T)^{3/4}}
    \left[F\left(\frac{(1+\alpha_T)^{1/4}}{M{\cal R}}\right)'\right]'}.
    \label{eq:Veff_general}
    \end{align}

As a consistency check, let us choose
    \begin{align}
    M^2={\rm const}, \qquad
    \Lambda=c=\tilde{\beta}=\alpha=\zeta=M_3^2=0,
    \end{align}
which admits the stealth Schwarzschild background
    \begin{align}
    A=B=1-\frac{r_{\rm H}}{r}, \qquad
    {\cal R}=r, \qquad
    {\cal X}=1,
    \end{align}
as discussed in Ref.~\cite{Mukohyama:2022skk}.
Indeed, it is immediate to verify that the tadpole cancellation conditions~\eqref{eq:tadpole_spherical_1} and \eqref{eq:tadpole_spherical_2}, as well as the consistency relation~\eqref{eq:consistency_spherical}, are satisfied for this choice.
In this case, Eq.~\eqref{eq:Veff_general} reduces to
    \begin{align}
    V_{\rm eff}(r)
    =\left(1-\frac{r_{\rm H}}{r}\right)
    \left[
    \frac{\ell(\ell+1)}{r^2}
    -\frac{3r_{\rm H}}{r^3}
    \right]\eqqcolon V_{\rm RW}(r),
    \label{eq:VRW}
    \end{align}
which is the standard Regge-Wheeler potential.

In the earlier analyses of Refs.~\cite{Mukohyama:2022skk,Mukohyama:2023xyf}, the assumption~${\cal X}=1$ was motivated by typical stealth BH solutions, for which $\bar{X}=-q^2$ (see, e.g., Refs.~\cite{Mukohyama:2005rw,Babichev:2013cya,Kobayashi:2014eva,Babichev:2016kdt,Babichev:2017guv,Babichev:2017lmw,BenAchour:2018dap,Motohashi:2019sen,Charmousis:2019vnf,Minamitsuji:2019shy,Takahashi:2020hso}).
The choice~${\cal R}=r$ amounts to adopting the areal-radius gauge, while $M^2={\rm const}$ corresponds to working in an ``almost Einstein frame,'' i.e., a frame in which the coefficient of the Einstein-Hilbert term is constant~\cite{Mukohyama:2024pqe}.
As shown above, relaxing these assumptions does not change the basic structure of the generalized Regge-Wheeler equation, but allows us to accommodate more general backgrounds with a radially varying scalar kinetic term as well as geometries for which the areal radius has a turning point, such as wormhole spacetimes.
Keeping $M^2(r)$ arbitrary is also useful when matter fields are taken into account, since transforming to an almost Einstein frame generally renders the matter sector nonminimally coupled to the scalar field if it is minimally coupled in the original frame.

\section{Inverse design of the EFT}\label{sec:inverse_design}

\subsection{General observations}\label{ssec:general}

Having derived the generalized Regge-Wheeler equation, we now turn to the inverse-design problem.
Rather than specifying the background and EFT coefficients and deriving the effective potential, we reverse the logic and ask what background and EFT functions reproduce a target effective potential.

From Eq.~\eqref{eq:Veff_general}, one immediately sees that the effective potential has the form
    \begin{align}
    V_{\rm eff}(r)=(\ell-1)(\ell+2)V_\Omega(r)+V_0(r),
    \label{eq:Veff_decomp}
    \end{align}
where $V_\Omega$ and $V_0$ are independent of $\ell$.
Using Eqs.~\eqref{eq:master_coords} and \eqref{eq:F_def}, these functions can be written as
    \begin{align}
    V_\Omega=\frac{{\cal X}A+\alpha_T}{{\cal X}{\cal R}^2}, \qquad
    V_0
    =\frac{M{\cal R}}{(1+\alpha_T)^{1/4}}F\left[F\left(\frac{(1+\alpha_T)^{1/4}}{M{\cal R}}\right)'\,\right]'.
    \label{eq:VOmega_V0}
    \end{align}
This structure constrains the $\ell$ dependence of a family of target effective potentials that can be reproduced by a single background and a single set of EFT coefficients.
Specifically, we consider a prescribed family of effective potentials~$U(r)$ of the form
    \begin{align}
    U(r)=(\ell-1)(\ell+2)U_\Omega(r)+U_0(r),
    \label{eq:U_decomp}
    \end{align}
where $U_\Omega$ and $U_0$ are independent of $\ell$.
For a single fixed multipole, the decomposition~\eqref{eq:U_decomp} is of course not unique; the restriction becomes nontrivial when the same EFT realization is required to reproduce target potentials for more than one value of $\ell$.
Within the choice of coordinates and master variable adopted here, we look for EFT realizations satisfying
    \begin{align}
    V_\Omega=U_\Omega, \qquad
    V_0=U_0.
    \label{eq:reconstruction}
    \end{align}
A realization of the prescribed potentials is obtained if background and EFT functions satisfying the above reconstruction equations can be found.
They must also satisfy the condition~$1-{\cal X}A\ge 0$ in Eq.~\eqref{eq:1-XA}, the tadpole cancellation conditions~\eqref{eq:tadpole_spherical_1} and \eqref{eq:tadpole_spherical_2}, the consistency relation~\eqref{eq:consistency_spherical}, the restriction~$M_3^2+\alpha=0$ imposed in Eq.~\eqref{eq:p4_zero}, and the stability conditions~\eqref{eq:odd_health}.\footnote{In addition, global regularity of the scalar field may impose further restrictions on the reconstructed background and EFT parameters, depending on the horizon structure, as illustrated in Sec.~\ref{ssec:PT}.}
The conditions~\eqref{eq:reconstruction} should not be regarded as necessary conditions for physical equivalence, since different choices of radial coordinate or master variable can lead to different representations of the same perturbation equation.
In what follows, we restrict ourselves to realizations satisfying Eq.~\eqref{eq:reconstruction}.

Details of the reconstruction procedure depend on the explicit form of the target potential.
In what follows, we illustrate the inverse-design procedure with several examples.
First, in Sec.~\ref{ssec:PT}, we consider two analytically tractable backgrounds for which the effective potential takes the exactly solvable P\"{o}schl-Teller form: one admitting a Schwarzschild-de Sitter-type static patch and the other describing a wormhole-type static patch.
In Sec.~\ref{ssec:RW}, we next turn to the Regge-Wheeler potential and characterize EFT realizations that reproduce the GR potential exactly.
Finally, in Sec.~\ref{ssec:parametrized_deformations}, we consider parametrized deformations of the Regge-Wheeler potential and construct explicit realizations of such phenomenological modifications.

\subsection{P\"{o}schl-Teller potential}\label{ssec:PT}

As a first example, we consider the case where the target potential, i.e., $U(r)$ above, is given by the P\"{o}schl-Teller potential
    \begin{align}
    V_{\rm PT}(r_*)=\frac{v_\ell}{\sigma^2}\,{\rm sech}^2\,\frac{r_*}{\sigma},
    \label{eq:VPT}
    \end{align}
where the tortoise coordinate ranges over $-\infty<r_*<\infty$.
Here, $\sigma>0$ is a constant with dimensions of length that sets the width of the potential, and $v_\ell$ is a dimensionless amplitude that may depend on $\ell$.
One advantage of choosing the P\"{o}schl-Teller potential as the target is that its quasinormal-mode spectrum is known analytically, so that, once the potential is realized exactly, the corresponding odd-parity quasinormal-mode frequencies can be read off immediately.
With the time dependence $\Psi\propto {\rm e}^{-{\rm i}\omega\tilde{t}}$, imposing the quasinormal-mode boundary conditions $\Psi\propto {\rm e}^{-{\rm i}\omega r_*}$ as $r_*\to-\infty$ and $\Psi\propto {\rm e}^{+{\rm i}\omega r_*}$ as $r_*\to+\infty$ yields~\cite{Ferrari:1984zz}
    \begin{align}
    \omega
    =\frac{1}{\sigma}
    \left[
    \pm\sqrt{v_\ell-\frac{1}{4}}
    -{\rm i}\left(n+\frac{1}{2}\right)
    \right],
    \qquad n=0,1,2,\cdots,
    \end{align}
where we have assumed $v_\ell>1/4$.
Since $\tilde{t}$ differs from the static time coordinate~$t$ only by a radial shift, the frequencies defined with respect to $\tilde{t}$ coincide with those defined with respect to $t$.

In order for the same EFT realization to reproduce Eq.~\eqref{eq:VPT} for different multipoles, the $\ell$ dependence of $v_\ell$ is constrained by the decomposition~\eqref{eq:U_decomp}.
A particularly simple choice is $v_\ell=(\ell-1)(\ell+2)$, which corresponds to
    \begin{align}
    U_\Omega(r)=\frac{1}{\sigma^2}\,{\rm sech}^2\,\frac{r_*(r)}{\sigma}, \qquad
    U_0(r)=0.
    \label{eq:PT_target}
    \end{align}
The reconstruction conditions~\eqref{eq:reconstruction} are then satisfied if
    \begin{align}
    \frac{{\cal X}A+\alpha_T}{{\cal X}{\cal R}^2}=U_\Omega, \qquad
    \frac{(1+\alpha_T)^{1/4}}{M{\cal R}}=\gamma,
    \label{eq:PT_reconstruction}
    \end{align}
where $\gamma>0$ is a constant.
Combining these conditions with Eqs.~\eqref{eq:master_coords} and \eqref{eq:F_def}, we obtain
    \begin{align}
    \frac{\D r_*}{\D r}
    =\sqrt{\frac{A}{B}}\frac{\gamma^2M^2}{U_\Omega}
    =\gamma^2\sigma^2M^2\sqrt{\frac{A}{B}}
    \cosh^2\frac{r_*}{\sigma},
    \label{eq:PT_tortoise}
    \end{align}
which provides a relation between $r_*$ and $r$.

In what follows, we study two realizations of the P\"{o}schl-Teller potential within the EFT framework.
The first admits, for an appropriate parameter range, a static patch resembling that of Schwarzschild-de Sitter, while the second describes a wormhole-type static patch.

\subsubsection{A Schwarzschild-de Sitter-type static patch}\label{sssec:PT_BH-dS}

Let us assume ${\cal R}'>0$ and choose the areal-radius gauge by setting
    \begin{align}
    {\cal R}=r,
    \end{align}
and impose the following conditions:
    \begin{align}
    A=B, \qquad
    M^2={\rm const}>0.
    \end{align}
We restrict ourselves to $r>0$, as $r$ coincides with the areal radius in this gauge.
Then, using Eqs.~\eqref{eq:p4_zero}, \eqref{eq:alphaT_def}, and the second equation in Eq.~\eqref{eq:PT_reconstruction}, we obtain
    \begin{align}
    \alpha=M^2\bra{1-\frac{L^4}{r^4}},
    \end{align}
where we have defined $L\coloneqq(\gamma M)^{-1}$.
Together with Eq.~\eqref{eq:p4_zero}, this gives $M^2+M_3^2=M^2L^4/r^4>0$, so that the stability conditions~\eqref{eq:odd_health} are satisfied.
Also, Eq.~\eqref{eq:PT_tortoise} gives
    \begin{align}
    r=r_0+\frac{L^2}{\sigma}\tanh\frac{r_*}{\sigma},
    \end{align}
where $r_0$ is an integration constant.
As $r_*$ varies from $-\infty$ to $+\infty$, $r$ varies from $r_-$ to $r_+$, where
    \begin{align}
    r_\pm\coloneqq r_0\pm\frac{L^2}{\sigma}.
    \end{align}
Since $F=\D r/\D r_*$ vanishes at $r=r_\pm$, these endpoints correspond to odd-mode characteristic horizons, but they need not coincide with horizons of the background metric.
Accordingly, the quasinormal-mode boundary conditions are imposed at $r=r_\pm$.
We require $r_->0$, i.e., $r_0>L^2/\sigma$, so that the entire interval lies in the region~$r>0$.
The first equation in Eq.~\eqref{eq:PT_reconstruction} then yields
    \begin{align}
    A=\frac{1-(r/L)^4}{\cal X}
    +\frac{r^2}{\sigma^2}
    -\frac{r^2(r-r_0)^2}{L^4}.
    \end{align}
Substituting these expressions into the remaining tadpole cancellation condition~\eqref{eq:tadpole_spherical_2} gives an ordinary differential equation for ${\cal X}$, whose solution is
    \begin{align}
    {\cal X}=\frac{r}{r_1},
    \end{align}
where $r_1>0$ is an integration constant.
Thus, ${\cal X}$ is finite and positive throughout the interval~$r_-<r<r_+$.
Moreover, the condition~$1-{\cal X}A\geq0$ in Eq.~\eqref{eq:1-XA} is satisfied throughout this interval provided that
    \begin{align}
    r_1\geq 2\left(r_0-\sqrt{r_0^2-\frac{L^4}{\sigma^2}}\right).
    \label{eq:PT_regular}
    \end{align}
The remaining EFT functions~$\Lambda$, $c$, $\tilde{\beta}$, and $\zeta$ can then be determined from Eqs.~\eqref{eq:tadpole_spherical_1} and \eqref{eq:consistency_spherical}.
For the strict inequality in Eq.~\eqref{eq:PT_regular}, one has $1-{\cal X}A>0$ throughout the interval, whereas the limiting case in which the inequality is saturated will be of particular interest below.
Note also that, in this limiting case, ${\cal Z}$ vanishes at one point in the interval, as shown below.
Regularity of $\zeta$ at this point requires the integration constant in Eq.~\eqref{eq:consistency_spherical} to vanish, and hence $\zeta=0$ throughout the interval.

Let us briefly discuss the background geometry described by this solution.
To exhibit a static patch resembling that of Schwarzschild-de Sitter, we focus on the parameter region~$r_->L$, for which $A(r_\pm)<0$.
A static region bounded by two background horizons is realized if $A$ becomes positive somewhere between the two endpoints.
For example, the simple sufficient condition~$A(r_0)>0$ amounts to
    \begin{align}
    r_1<\frac{L^4r_0^3}{\sigma^2(r_0^4-L^4)}.
    \label{eq:PT_static}
    \end{align}
This condition is sufficient but not necessary for the existence of such a static patch, and it is compatible with the regularity condition~\eqref{eq:PT_regular} in a nonempty region of parameter space.
The spacetime should not, however, be regarded as asymptotically de Sitter, since its extension outside the region of interest generally has a different global structure.

We finally discuss the regularity of the scalar field at the two future horizons of the background metric.
A scalar field with a fixed branch choice for $\psi'$ cannot in general be regular at both the future BH and future cosmological horizons~\cite{Kobayashi:2014eva,Takahashi:2020hso}.
Indeed, Eq.~\eqref{eq:1-XA} gives
    \begin{align}
    \psi'=\pm q\sqrt{\frac{1-{\cal X}A}{AB}}.
    \end{align}
To see the required branch choice explicitly, let us introduce the ingoing and outgoing Eddington-Finkelstein coordinates associated with the background metric,
    \begin{align}
    \D v=\D t+\frac{\D r}{\sqrt{AB}}, \qquad
    \D u=\D t-\frac{\D r}{\sqrt{AB}}.
    \end{align}
Since $1-{\cal X}A\to1$ at a horizon, the plus and minus branches behave as $\bar\phi\simeq qv$ and $\bar\phi\simeq qu$, respectively, up to terms regular at the horizon, where we have used $\bar\phi=qt+\psi$.
Hence, regularity at the future BH horizon requires the plus branch, whereas regularity at the future cosmological horizon requires the minus branch~\cite{Kobayashi:2014eva,Takahashi:2020hso}.
Therefore, a scalar field regular at both future horizons must allow $\psi'$ to change sign somewhere inside the static region.

Such a smooth sign change is possible when $r_1$ is chosen so that Eq.~\eqref{eq:PT_regular} is saturated, in analogy with the mechanism discussed in Ref.~\cite{Charmousis:2019vnf}.
At this value, one finds
    \begin{align}
    1-{\cal X}A
    =\frac{r^3}{r_1L^4}\bra{r-r_{\rm tp}}^2, \qquad
    r_{\rm tp}\coloneqq \sqrt{r_0^2-\frac{L^4}{\sigma^2}},
    \label{eq:PT_scalar_double_zero}
    \end{align}
so that $1-{\cal X}A$ has a double zero at $r=r_{\rm tp}$.
As shown below, $r_{\rm tp}$ corresponds to the turning point of the radial scalar profile, where $\psi'$ changes sign.
In particular, since $0<r_-<r_{\rm tp}<r_0<r_+$, $A(r_\pm)<0$, and $A(r_{\rm tp})=r_1/r_{\rm tp}>0$, continuity of $A$ guarantees the existence of at least one background Killing horizon on each side of $r_{\rm tp}$.
Thus, the turning point lies in a static region bounded by background horizons.
A smooth choice of the scalar profile is then obtained by choosing
    \begin{align}
    \psi'
    =\frac{q}{A}\frac{r^{3/2}(r_{\rm tp}-r)}{\sqrt{r_1}L^2},
    \label{eq:PT_scalar_signed}
    \end{align}
which corresponds to switching between the two branches of $\psi'$ at $r=r_{\rm tp}$.
Thus, $\psi'$ vanishes at $r=r_{\rm tp}$ and changes smoothly from positive to negative as $r$ increases through $r_{\rm tp}$, so that the scalar field is regular at both future horizons.

\subsubsection{A wormhole-type static patch}\label{sssec:PT_wormhole}

Let us adopt the conformal gauge for the two-dimensional $(t,r)$ sector by setting
    \begin{align}
    B=\frac{1}{A},
    \end{align}
and impose
    \begin{align}
    \alpha=M_3^2=0.
    \end{align}
Then, $\alpha_T=0$, and Eq.~\eqref{eq:F_def} gives $F=1$, so that we can choose $r_*=r$.
We further consider the following ansatz for the areal radius:
    \begin{align}
    {\cal R}={\cal R}_{\rm th}
    +({\cal R}_\infty-{\cal R}_{\rm th})
    \tanh^2\frac{r}{\sigma},
    \end{align}
where ${\cal R}_{\rm th}$ and ${\cal R}_\infty$ are positive constants satisfying ${\cal R}_{\rm th}<{\cal R}_\infty$.
The areal radius has a minimum~${\cal R}={\cal R}_{\rm th}$ at $r=0$ and approaches ${\cal R}_\infty$ as $r\to\pm\infty$.

The second equation of Eq.~\eqref{eq:PT_reconstruction} yields
    \begin{align}
    M^2=\frac{1}{\gamma^2{\cal R}^2},
    \end{align}
which is positive and hence, together with $M_3^2=0$, ensures the stability conditions~\eqref{eq:odd_health}, while the first equation gives
    \begin{align}
    A=\frac{{\cal R}^2}{\sigma^2}\,{\rm sech}^2\,\frac{r}{\sigma}.
    \end{align}
Since $A(0)={\cal R}_{\rm th}^2/\sigma^2>0$ and ${\cal R}(0)={\cal R}_{\rm th}$ is a regular minimum, the surface~$r=0$ represents a regular wormhole throat.
Moreover, as $r\to\pm\infty$, $A$ vanishes while ${\cal R}$ approaches the finite value~${\cal R}_\infty$, and each endpoint corresponds to a Killing horizon with a near-horizon structure analogous to that of the cosmological horizon of a de Sitter static patch.
Hence, the throat lies in a wormhole-type static patch bounded by two Killing horizons, both having areal radius~${\cal R}_\infty$.
In this case, the odd-mode characteristic horizons coincide with these background Killing horizons, so that the quasinormal-mode boundary conditions are imposed at these two horizons.
Substituting these expressions into the tadpole cancellation condition~\eqref{eq:tadpole_spherical_2} gives an ordinary differential equation for ${\cal X}$, whose solution is
    \begin{align}
    {\cal X}=\frac{{\cal X}_0{\cal R}_{\rm th}^3}{{\cal R}^3},
    \end{align}
where ${\cal X}_0>0$ is an integration constant.
The condition~$1-{\cal X}A\ge 0$ in Eq.~\eqref{eq:1-XA} is satisfied throughout the static region provided that
    \begin{align}
    {\cal X}_0\le \frac{\sigma^2}{{\cal R}_{\rm th}^2}.
    \label{eq:PT_wormhole_regular}
    \end{align}
Since ${\cal X}'$ vanishes at the throat while the other factors in Eq.~\eqref{eq:consistency_spherical} remain finite there, regularity of $\zeta$ requires the integration constant in the consistency relation to vanish.
Since ${\cal X}'\ne0$ away from the throat at finite $r$, this fixes $\zeta=0$ throughout the static patch.
The remaining EFT functions~$\Lambda$, $c$, and $\tilde{\beta}$ are determined by Eq.~\eqref{eq:tadpole_spherical_1} and remain regular throughout the static patch.

Finally, let us briefly discuss the regularity of the scalar field at the two future horizons.
As in the Schwarzschild-de Sitter-type case, regularity at both horizons requires $\psi'$ to change sign between them.
This can be achieved by saturating Eq.~\eqref{eq:PT_wormhole_regular}, namely, by setting ${\cal X}_0=\sigma^2/{\cal R}_{\rm th}^2$.
In this case,
    \begin{align}
    1-{\cal X}A
    =\frac{{\cal R}_\infty}{\cal R}\tanh^2\frac{r}{\sigma},
    \end{align}
and a smooth choice of $\psi'$ is given by
    \begin{align}
    \psi'
    =-q\sqrt{\frac{{\cal R}_\infty}{\cal R}}
    \tanh\frac{r}{\sigma}.
    \end{align}
Thus, $\psi'$ vanishes smoothly at the throat and approaches $+q$ and $-q$ as $r\to-\infty$ and $r\to+\infty$, respectively.
Since $AB=1$, this branch behaves as $\bar\phi\simeq qv$ and $\bar\phi\simeq qu$ at the left and right future horizons, respectively, where $v=t+r$ and $u=t-r$ are the corresponding null coordinates of the background metric.
Hence, the scalar field is regular at both future horizons.

\subsection{Regge-Wheeler potential}\label{ssec:RW}

Let us consider the case where the target potential is the Regge-Wheeler potential~\eqref{eq:VRW}, which amounts to
    \begin{align}
    U_\Omega(r)=\frac{f}{r^2}, \qquad
    U_0(r)=f\frac{2r-3r_{\rm H}}{r^3}, \qquad
    f(r)\coloneqq 1-\frac{r_{\rm H}}{r}.
    \label{eq:RW_target}
    \end{align}
For simplicity, we assume $F=f$ so that $r_*$ coincides with the Schwarzschild tortoise coordinate.
From Eq.~\eqref{eq:F_def}, this implies
    \begin{align}
    \sqrt{\frac{B}{A}}\frac{{\cal X}A+\alpha_T}{{\cal X}\sqrt{1+\alpha_T}}=f.
    \label{eq:F=f}
    \end{align}
The reconstruction conditions~\eqref{eq:reconstruction} are then satisfied if
    \begin{align}
    \frac{{\cal X}A+\alpha_T}{{\cal X}{\cal R}^2}=\frac{f}{r^2}, \qquad
    \frac{M{\cal R}}{(1+\alpha_T)^{1/4}}\left[f\left(\frac{(1+\alpha_T)^{1/4}}{M{\cal R}}\right)'\,\right]'=\frac{2r-3r_{\rm H}}{r^3}.
    \label{eq:RW_reconstruction}
    \end{align}
These equations yield
    \begin{align}
    A=\frac{{\cal R}^4}{r^4(1+\alpha_T)}B=\frac{{\cal R}^2}{r^2}f-\frac{\alpha_T}{\cal X}, \qquad
    \frac{M{\cal R}}{(1+\alpha_T)^{1/4}}=M_0 r,
    \label{eq:RW_ABM}
    \end{align}
with $M_0>0$ an integration constant.
Here, we have selected the solution regular at $r=r_{\rm H}$; the second independent solution for $(1+\alpha_T)^{1/4}/(M{\cal R})$ diverges logarithmically there.
The above equations, together with Eqs.~\eqref{eq:p4_zero} and \eqref{eq:alphaT_def}, yield
    \begin{align}
    \alpha=-M_3^2=\frac{r^2\alpha_T}{{\cal R}^2\sqrt{1+\alpha_T}}M_0^2.
    \end{align}
For $1+\alpha_T>0$, the right-hand side is a monotonic function of $\alpha_T$, so this relation uniquely determines $\alpha_T$ for given $\alpha$ and ${\cal R}$.
Substituting the above expressions into the tadpole cancellation condition~\eqref{eq:tadpole_spherical_2}, we find
    \begin{align}
    \bra{\frac{\cal R}{r}}'\brb{\frac{{\cal R}^3{\cal X}}{r^3(1+\alpha_T)}}'=0.
    \label{eq:tadpole_alpha_RW}
    \end{align}

This equation admits two branches.
The first branch is characterized by ${\cal R}/r={\rm const}$.
For simplicity, we set this constant to unity, ${\cal R}=r$.
In this case, the functions~${\cal X}$ and $\alpha$ remain arbitrary, as long as the condition~$1-{\cal X}A\ge 0$ in Eq.~\eqref{eq:1-XA} is satisfied.
The stability conditions~\eqref{eq:odd_health} can equivalently be written as $M^2>0$ and $1+\alpha_T>0$, and are therefore automatically satisfied in the present construction.
Once ${\cal X}$ and $\alpha$ are specified, $A$, $B$, $M^2$, and $M_3^2$ are fixed by the above relations, while the remaining EFT coefficients can be obtained from Eqs.~\eqref{eq:tadpole_spherical_1} and \eqref{eq:consistency_spherical}.
Thus, there is a large family of EFT realizations of the Regge-Wheeler potential.
For ${\cal X}=1$ and $\alpha=0$, the Schwarzschild background~$A=B=f$ is recovered.
Conversely, requiring the background metric to be Schwarzschild fixes $\alpha=0$, while ${\cal X}$ remains arbitrary.
For a related discussion of the restrictions on the effective potential when the background geometry and the generalized tortoise coordinate are fixed to their Schwarzschild forms, see the \hyperref[app:fixed_Schwarzschild]{Appendix}.

The second branch is characterized by
    \begin{align}
    \frac{{\cal R}^3{\cal X}}{r^3(1+\alpha_T)}={\cal C},
    \label{eq:branch2}
    \end{align}
with ${\cal C}>0$ an integration constant.
Unlike the first branch, this relation imposes a nontrivial relation among ${\cal R}$, ${\cal X}$, and $\alpha_T$.
Given ${\cal R}$ and $\alpha$, the function~$\alpha_T$ is fixed by the algebraic relation above, while ${\cal X}$ is then determined by this equation up to a constant.
Thus, ${\cal R}$ and $\alpha$ may still be chosen freely, subject to the condition~$1-{\cal X}A\ge 0$, while the other functions are determined by the relations above and the remaining tadpole and consistency conditions.

\subsection{Parametrized deformations of the Regge-Wheeler potential}\label{ssec:parametrized_deformations}

We now consider parametrized deformations of the Regge-Wheeler potential.
Specifically, we consider a target potential with power-law corrections of the form
    \begin{align}
    U(r)
    =V_{\rm RW}(r)
    +\frac{f}{r_{\rm H}^2}
    \sum_j\alpha_j
    \bra{\frac{r_{\rm H}}{r}}^j,
    \label{eq:pRW_full_coeffs}
    \end{align}
where the dimensionless coefficients~$\alpha_j$ may depend on $\ell$ and are assumed to be sufficiently small.
Here and in what follows, $j$ runs over nonnegative integers unless otherwise specified.
If only a finite number of power-law corrections are included, all other coefficients are understood to vanish.
For the same background and EFT coefficients to reproduce such a family for different multipoles, the decomposition~\eqref{eq:Veff_decomp} requires the $\ell$ dependence of $\alpha_j$ to take the form
    \begin{align}
    \alpha_j
    =(\ell-1)(\ell+2)\alpha_j^{(\Omega)}
    +\alpha_j^{(0)},
    \label{eq:pRW_multipole_filter}
    \end{align}
where $\alpha_j^{(\Omega)}$ and $\alpha_j^{(0)}$ are independent of $\ell$.
Thus, once $\alpha_j$ is specified for any two multipoles, its values for all other multipoles are fixed, so that nontrivial relations arise among three or more multipoles.
Accordingly, we write
    \begin{align}
    U_\Omega(r)
    &=f\bra{\frac{1}{r^2}+\delta U_\Omega}, \qquad
    U_0(r)
    =f\bra{\frac{2r-3r_{\rm H}}{r^3}+\delta U_0},
    \label{eq:U_Omega,0}
    \end{align}
with
    \begin{align}
    \delta U_\Omega
    &=\frac{1}{r_{\rm H}^2}\sum_j\alpha_j^{(\Omega)}
    \bra{\frac{r_{\rm H}}{r}}^j, \qquad
    \delta U_0
    =\frac{1}{r_{\rm H}^2}\sum_j\alpha_j^{(0)}
    \bra{\frac{r_{\rm H}}{r}}^j.
    \label{eq:deltaU_Omega,0}
    \end{align}
Such power-law parametrizations of small deviations from the Regge-Wheeler potential have been used in various contexts, including BH quasinormal modes, tidal Love numbers, and tidal dissipation numbers (see, e.g., Refs.~\cite{Cardoso:2019mqo,McManus:2019ulj,Katagiri:2023umb,Barura:2024uog,Kobayashi:2025swn}).
The inverse problem of reconstructing modifications of the effective potential from quasinormal-mode spectra has also been studied~\cite{Volkel:2022khh}.
Our aim here is complementary: we construct an explicit EFT realization that gives rise to a parametrically deformed Regge-Wheeler potential of the above form.

As we are interested in perturbative deformations away from the Regge-Wheeler case, we seek solutions perturbatively around the Regge-Wheeler realizations found in Sec.~\ref{ssec:RW}.
Again, we assume $F=f$, so that Eq.~\eqref{eq:F=f} holds.
The reconstruction conditions~\eqref{eq:reconstruction} are then satisfied if
    \begin{align}
    \frac{{\cal X}A+\alpha_T}{{\cal X}{\cal R}^2}=f\bra{\frac{1}{r^2}+\delta U_\Omega}, \qquad
    \frac{M{\cal R}}{(1+\alpha_T)^{1/4}}\left[f\left(\frac{(1+\alpha_T)^{1/4}}{M{\cal R}}\right)'\,\right]'=\frac{2r-3r_{\rm H}}{r^3}+\delta U_0.
    \label{eq:pRW_reconstruction}
    \end{align}
From the first equation and Eq.~\eqref{eq:F=f}, to first order in the deformations, we have
    \begin{align}
    A={\cal R}^2f\bra{\frac{1}{r^2}+\delta U_\Omega}-\frac{\alpha_T}{\cal X}, \qquad
    B=\frac{r^4(1+\alpha_T)}{{\cal R}^4}(1-2r^2\delta U_\Omega)A.
    \end{align}
Then, solving the second equation in Eq.~\eqref{eq:pRW_reconstruction} and choosing the solution regular at the horizon, we find
    \begin{align}
    \frac{M{\cal R}}{(1+\alpha_T)^{1/4}}=M_0 r\brb{1+\gamma_1(r)}, \qquad
    \gamma_1'=-\frac{r^2}{f}\int_{r_{\rm H}}^r\D \tilde r\frac{\delta U_0(\tilde r)}{\tilde r^2}.
    \end{align}
The additive integration constant in $\gamma_1$ can be absorbed into a redefinition of $M_0$.
The lower integration limit in the expression for $\gamma_1'$ ensures that $\gamma_1'$ is regular at the horizon, provided that $\delta U_0$ is regular there.
The above equations, together with Eqs.~\eqref{eq:p4_zero} and \eqref{eq:alphaT_def}, yield
    \begin{align}
    \alpha=-M_3^2=\frac{r^2\alpha_T}{{\cal R}^2\sqrt{1+\alpha_T}}\brb{1+2\gamma_1(r)}M_0^2.
    \end{align}
To perturb around the branch~\eqref{eq:branch2} of the Regge-Wheeler case, we write
    \begin{align}
    \frac{{\cal R}^3{\cal X}}{r^3(1+\alpha_T)}={\cal C}\brb{1+\gamma_2(r)},
    \end{align}
where ${\cal C}$ is a constant and $\gamma_2$ is of first order in the deformations.
Substituting the above expressions into the tadpole cancellation condition~\eqref{eq:tadpole_spherical_2}, we then find
    \begin{align}
    \bra{\frac{\cal R}{r}}'\gamma_2'
    =-\frac{{\cal R}^4}{r^4}\bra{\frac{r^3}{{\cal R}^3}\gamma_1'}'
    +\frac{\cal R}{r^2}\bra{r^2\delta U_\Omega}'
    -{\cal C}\brc{\frac{r^2}{2}\bra{\frac{f}{r^2}}'\gamma_1'+\frac{1}{2}\delta U_\Omega+\frac{1}{4r^2}\brb{r^2f\bra{r^2\delta U_\Omega}'}'}.
    \end{align}
The remaining EFT coefficients~$\Lambda$, $c$, $\tilde{\beta}$, and $\zeta$ can then be obtained from Eqs.~\eqref{eq:tadpole_spherical_1} and \eqref{eq:consistency_spherical}.

As a simple explicit demonstration, let us choose $\alpha=M_3^2=0$ and
    \begin{align}
    {\cal R}=r\brb{1+\lambda\bra{\frac{r_{\rm H}}{r}}^2},
    \end{align}
with $0<\lambda<1$.
This choice gives $\alpha_T=0$ and ${\cal R}'>0$ for $r\geq r_{\rm H}$.
For a single-power deformation of $\delta U_\Omega$,
    \begin{align}
    \delta U_\Omega=\frac{\alpha_j^{(\Omega)}}{r_{\rm H}^2}\bra{\frac{r_{\rm H}}{r}}^j, \qquad
    \delta U_0=0,
    \end{align}
with $j\geq5$, we have $\gamma_1=0$, and the reconstructed background and EFT functions are
    \begin{align}
    \begin{split}
    &A=f\bra{\frac{\cal R}{r}}^2
    \brb{1+\alpha_j^{(\Omega)}\bra{\frac{r_{\rm H}}{r}}^{j-2}}, \qquad
    B=f\bra{\frac{r}{\cal R}}^2
    \brb{1-\alpha_j^{(\Omega)}\bra{\frac{r_{\rm H}}{r}}^{j-2}}, \\
    &M^2=M_0^2\bra{\frac{r}{\cal R}}^2, \qquad
    {\cal X}
    ={\cal C}\bra{\frac{r}{\cal R}}^3
    \brb{1+\gamma_2(r)},
    \end{split}
    \end{align}
where, to first order in $\alpha_j^{(\Omega)}$,
    \begin{align}
    \gamma_2
    =-\frac{\alpha_j^{(\Omega)}}{2\lambda}
    \biggl[
    &\frac{4(j-2)+{\cal C}(j^2-5j+8)}{4(j-4)}
    \bra{\frac{r_{\rm H}}{r}}^{j-4}
    -\frac{{\cal C}(j-2)^2}{4(j-3)}
    \bra{\frac{r_{\rm H}}{r}}^{j-3}
    +\lambda\bra{\frac{r_{\rm H}}{r}}^{j-2}
    \biggr].
    \end{align}
Here, the additive integration constant in $\gamma_2$ has been absorbed into ${\cal C}$.
For $0<{\cal C}<1$ and sufficiently small  $|\alpha_j^{(\Omega)}|/\lambda$, the condition~$1-{\cal X}A\ge 0$ is satisfied, while $M^2>0$ and $M_3^2=0$ ensure the stability conditions~\eqref{eq:odd_health}.
Note that the limit~$\lambda\to0^+$ at fixed nonzero $\alpha_j^{(\Omega)}$ is not covered by this perturbative solution.
The extension to a superposition of several powers is immediate at linear order.

By contrast, a single-power deformation of $\delta U_0$,
    \begin{align}
    \delta U_\Omega=0, \qquad
    \delta U_0=\frac{\alpha_j^{(0)}}{r_{\rm H}^2}\bra{\frac{r_{\rm H}}{r}}^j,
    \end{align}
generally leads to a divergent $\gamma_1$ at spatial infinity,
    \begin{align}
    r_{\rm H}\gamma_1'=-\frac{\alpha_j^{(0)}}{j+1}\frac{r^2}{r_{\rm H}^2f}+{\cal O}\bra{\bra{\frac{r_{\rm H}}{r}}^{j-1}}.
    \end{align}
For a superposition of several powers, however, this leading divergent behavior can be canceled by imposing
    \begin{align}
    \sum_j\frac{\alpha_j^{(0)}}{j+1}=0.
    \end{align}
If all nonvanishing powers satisfy $j\geq3$, $\gamma_1$ then approaches a constant at spatial infinity, which can be absorbed into a redefinition of $M_0$, so that we may choose $\gamma_1\to0$ as $r\to\infty$.

The explicit realization above need not be close to the Schwarzschild geometry at the level of the background metric.
Consequently, observables governed by minimally coupled matter, such as photon trajectories, can in general deviate from their GR counterparts independently of the modification of the odd-parity potential.
If such observables are required to remain close to their GR predictions, it is natural to seek realizations for which the background metric itself is perturbatively close to Schwarzschild, e.g., ${\cal R}=r$ and $A,B=f+{\cal O}(\alpha_j^{(\Omega)},\alpha_j^{(0)})$.
Such a reconstruction, however, is more subtle.
In the exact Regge-Wheeler limit, Eq.~\eqref{eq:tadpole_alpha_RW} shows that the tadpole cancellation condition becomes degenerate on the ${\cal R}=r$ branch, where the first factor vanishes identically, while the standard stealth configuration with ${\cal X}/(1+\alpha_T)={\rm const}$ also makes the second factor vanish and therefore lies at the intersection of the two branches.
A perturbative reconstruction around such a degenerate configuration therefore requires a separate analysis, which we leave for future work.
For comparison, a more restricted setup in which the background geometry and the generalized tortoise coordinate are fixed to their Schwarzschild forms is considered in the \hyperref[app:fixed_Schwarzschild]{Appendix}.

\section{Conclusions}\label{sec:conc}

In this paper, we have investigated how a prescribed odd-parity effective potential can be realized within the EFT of perturbations in scalar-tensor gravity around a static and spherically symmetric background.
To accommodate a wider variety of such backgrounds, in Sec.~\ref{sec:model}, we have also extended the previous odd-parity EFT analysis by relaxing several assumptions imposed in earlier studies.
In particular, we have allowed for a radially varying scalar kinetic term and effective Planck mass, as well as a general areal-radius function.
The resulting generalized Regge-Wheeler equation involves the effective potential given by Eq.~\eqref{eq:Veff_general}, which encodes information about the background and EFT parameters.
We have then formulated the inverse-design problem in Sec.~\ref{ssec:general}.
A key observation is that the effective potential has the specific $\ell$ dependence shown in Eq.~\eqref{eq:Veff_decomp}, which constrains the form of target odd-parity potentials that can be realized within the EFT.
As analytically tractable examples, in Sec.~\ref{ssec:PT}, we have constructed two explicit realizations of the P\"{o}schl-Teller potential: one admitting a Schwarzschild-de Sitter-type static patch and the other describing a wormhole-type static patch.
We have then considered the Regge-Wheeler potential in Sec.~\ref{ssec:RW} and shown that it admits a large family of EFT realizations.
Thus, even exact agreement of the odd-parity effective potential, and hence of its quasinormal-mode spectrum, with the GR result does not uniquely determine the background geometry or the EFT data.
Finally, in Sec.~\ref{ssec:parametrized_deformations}, we have considered parametrized deformations of the standard Regge-Wheeler potential and constructed explicit EFT realizations, providing a route to interpreting phenomenological constraints on potential deformations from BH ringdown observations in terms of the underlying EFT data.

Several directions remain for future work.
In the present paper, we have focused exclusively on the odd-parity sector, and it would be important to extend the analysis to the even-parity sector and investigate the stability of complete EFT realizations.
In particular, for realizations of the standard Regge-Wheeler potential with an additional propagating scalar degree of freedom, the usual odd-even isospectrality of GR is generally expected to be broken, and it would be interesting to investigate this explicitly.
Another important direction is to study whether and how the reconstructed EFT data can be embedded into covariant scalar-tensor theories.
Such a covariant completion would provide a further connection between the perturbation dynamics and the underlying gravitational theory.


\acknowledgments{
K.T.~was supported in part by JSPS (Japan Society for the Promotion of Science) KAKENHI Grant No.~JP23K13101.
}


\appendix*
\section{Odd-parity effective potential on a fixed Schwarzschild background}
\label{app:fixed_Schwarzschild}

In the main text, we have treated the background geometry and the EFT coefficients simultaneously as quantities to be reconstructed from a prescribed effective potential.
For comparison, in this Appendix, we consider a more restricted setup in which the background geometry and the generalized tortoise coordinate are fixed in advance to their Schwarzschild forms,
    \begin{align}
    A=B=f=1-\frac{r_{\rm H}}{r}, \qquad
    {\cal R}=r, \qquad
    F=f,
    \label{eq:app_Schwarzschild_setup}
    \end{align}
for $r>r_{\rm H}$.
If matter is minimally coupled to $g_{\mu\nu}$, test-particle motion is therefore identical to that in Schwarzschild spacetime, while deviations from GR may still arise at the level of perturbations.

Let us clarify what form of the odd-parity effective potential is allowed in this setup.
We impose $1-{\cal X}f\geq0$, as required by Eq.~\eqref{eq:1-XA}, together with $1+\alpha_T>0$, which follows from Eqs.~\eqref{eq:odd_health} and \eqref{eq:alphaT_def}.
With these conditions, Eq.~\eqref{eq:F_def} under the restriction~\eqref{eq:app_Schwarzschild_setup} implies $\alpha_T=0$.
It then follows from Eqs.~\eqref{eq:p4_zero} and \eqref{eq:alphaT_def} that
    \begin{align}
    M_3^2=\alpha=0.
    \end{align}
Consequently, Eq.~\eqref{eq:VOmega_V0} reduces to
    \begin{align}
    V_\Omega=\frac{f}{r^2}, \qquad
    V_0
    =Mrf\brb{f\bra{\frac{1}{Mr}}'}'.
    \label{eq:app_VOmega_V0}
    \end{align}
Thus, within the restricted setup~\eqref{eq:app_Schwarzschild_setup}, the $\ell$-dependent part of the odd-parity effective potential is necessarily fixed to its Regge-Wheeler form, while possible deviations from GR can only reside in the $\ell$-independent part~$V_0$.

The tadpole cancellation condition~\eqref{eq:tadpole_spherical_2} also reduces to
    \begin{align}
    \brb{\frac{1}{\cal X}(M^2)'}'
    +\frac{1-3f}{2r}(M^2)'=0,
    \label{eq:app_alpha_tadpole}
    \end{align}
which provides an additional relation between $M^2$ and ${\cal X}$.
For the exact Regge-Wheeler potential, the regular branch selected in Sec.~\ref{ssec:RW} reduces to $M=M_0$ in the present setup.
In this case, Eq.~\eqref{eq:app_alpha_tadpole} is identically satisfied, leaving ${\cal X}$ arbitrary subject to $1-{\cal X}f\geq0$.
Thus, even after fixing the background geometry to Schwarzschild, the exact Regge-Wheeler limit retains a degeneracy in the scalar kinetic profile.

This restricted setup therefore illustrates how fixing the background geometry and the generalized tortoise coordinate substantially reduces the freedom of the inverse-design problem.
In particular, any effective potential realizable within Eq.~\eqref{eq:app_Schwarzschild_setup} must have $V_\Omega=f/r^2$, so that possible deviations from the Regge-Wheeler potential are confined to the $\ell$-independent part~$V_0$ and are further subject to the condition~\eqref{eq:app_alpha_tadpole}.


\bibliographystyle{mybibstyle}
\bibliography{bib}

\end{document}